\documentclass[useAMS,usenatbib, graphics]{mn2e}
\usepackage{txfonts}
\usepackage{graphicx}
\usepackage{hyperref} 
\usepackage{color}
\newcommand{\Teff}{$T_{\rm eff}$}
\newcommand{\apj}{ApJ}
\newcommand{\aap}{A\&A}

\newcommand{\aj}{AJ}

\newcommand{\apjl}{ApJL}
\newcommand{\mnras}{MNRAS}
\newcommand{\aapr}{A\&ARv}
\newcommand{\aaps}{Astronomy and Astrophysics Supplement}

\newcommand{\nat}{Nature}
\long\def\symbolfootnote[#1]#2{\begingroup%
\def\thefootnote{\fnsymbol{footnote}}\footnote[#1]{#2}\endgroup}
\title[LITHIUM IN GC GIANTS]{Lithium abundances in globular cluster giants: NGC 1904, NGC 2808, and NGC 362\thanks{Based on observations taken with ESO telescopes under program 094.D-0363(A)}}
\author[V. D'Orazi et al.]{V. D'Orazi$^{1,2,3}$, R. G. Gratton$^{1}$, G. C. Angelou$^{4,5}$, A. Bragaglia$^{6}$, E. Carretta$^{6}$, 
\newauthor J. C. Lattanzio$^{3}$, S. Lucatello$^{1}$, Y. Momany$^{1,7}$, A. Sollima$^{6}$, and G. Beccari$^{7,8}$\\
$^{1}$INAF- Osservatorio Astronomico di Padova, Vicolo dell'Osservatorio 5, 35122, Padova, Italy\\
$^{2}$Department of Physics and Astronomy, Macquarie University, Sydney, NSW 2109, Australia.\\
$^{3}$Monash Centre for Astrophysics,  School of Mathematical Sciences, Monash University,  Melbourne,  VIC 3800,  Australia.\\
$^{4}$Max Planck Institut f\"{u}r Sonnensystemforschung, Justus-von-Liebig-Weg 3, 37077 G\"{o}ttingen, Germany\\
$^{5}$Stellar Astrophysics Centre, Department of Physics and Astronomy, Aarhus University, Ny Munkegade 120, 8000 Aarhus C, Denmark\\
$^{6}$INAF - Osservatorio Astronomico di Bologna, via Ranzani 1, 40127, Bologna, Italy\\ 
$^{7}$European Southern Observatory, Alonso de Cordova 3107, Vitacura, Santiago, Chile\\
$^{8}$European Southern Observatory, Karl-Schwarzschild-Stra{\ss}e 2, 85748, Garching bei Munchen, Germany}

\begin{document}

\date{Accepted, 17 March 2015. Received, 17 March 2015; in original form, 10 March 2015}

\pagerange{\pageref{firstpage}--\pageref{lastpage}} \pubyear{2015}

\maketitle

\label{firstpage}

\begin{abstract}
The presence of multiple populations in globular clusters has been well established thanks to 
high-resolution spectroscopy.  It is widely accepted that distinct populations
are a consequence of different stellar generations: intra-cluster pollution episodes are required to produce the peculiar chemistry observed in almost all clusters. Unfortunately, the progenitors responsible have left an ambiguous signature and their
nature remains unresolved.  To constrain the candidate polluters, we have measured lithium and aluminium abundances in more than 180 giants across three systems: NGC~1904, NGC~2808, and NGC~362. The present investigation along with our previous analysis of M12 and M5 affords us the largest database of simultaneous determinations of Li and Al abundances.  Our results indicate that Li production has occurred in each of the three clusters.
In NGC~362 we  detected an M12-like behaviour, with first and second-generation stars sharing very similar Li abundances favouring a progenitor that is able to produce Li, such as AGB stars. Multiple progenitor types are possible in NGC~1904 and NGC~2808, as they possess both an intermediate population comparable in lithium to the first generation stars and also an extreme population, that is enriched in Al but depleted in Li. A simple dilution model fails in reproducing this complex pattern. Finally, the internal Li variation seems to suggest that the production efficiency of this element is a
function of the cluster's mass and metallicity - low-mass or relatively metal-rich clusters are more adept at producing Li.
\end{abstract}


\section{Introduction}\label{sec:intro}

Globular clusters (GCs) display internal variations in elements affected by proton-capture reactions (e.g., C, N, O, F, Na, Mg, and Al; \citealt{gratton12}) and exhibit split evolutionary sequences (\citealt{piotto12}; \citealt{milone12}) in their colour-magnitude diagrams. The presence of both features requires that GCs comprise multiple populations and it is this revelation that has renewed interest in these old stellar systems. In order to decipher their complex chemical history,  a concerted and systematic effort has been made to survey GCs for key diagnostic chemical species. Abundances have been determined for a considerable number of stars, from the main sequence through to the asymptotic giant branches (AGB), via different approaches and techniques (e.g., \citealt{carretta06}, \citeyear{carretta07}; \citealt{jp10}; \citealt{gratton11}, \citeyear{gratton13}, \citeyear{gratton14}; \citealt{campbell13}; \citealt{marino11}, \citeyear{marino13}, \citeyear{marino14}; \citealt{yong13}, \citeyear{yong14}, \citeyear{yong15}; \citealt{johnson15}; \citealt{lamb15}; \citealt{valenti15}; \citealt{cordero15}; \citealt{mucciarelli15}, and references therein). 

Red giant branches have been extensively studied with the largest survey of chemical compositions for GC giants 
undertaken by Carretta and collaborators  
(\citealt{carretta09a}, \citeyear{carretta09b}, \citeyear{carretta11}, \citeyear{carretta13}, \citeyear{carretta14a}).
They determined iron-peak, $\alpha$-, proton-capture, and neutron-capture element abundances for thousands of giants in 25 GCs, characterised by different morphological properties (i.e., mass, metallicity, age, environment, concentration, horizontal branch morphology). Recently, the northern sky has been surveyed by the SDSS-III Apache Point Observatory Galactic Evolution Experiment (APOGEE) and \cite{meszaros15} provided light element abundances for approximately 400 giants across ten GCs.

To date, almost every system (see however the peculiar cases of Ruprecht~106, Terzan 7, and Pal 12; \citealt{villanova13}; \citealt{sbordone07}; \citealt{cohen04}) analysed displays evidence for multiple stellar populations, suggesting that this feature is in fact ubiquitous. Spectroscopically this evidence is in the form of the characteristic GC abundance patterns which are present in the constituent dwarfs and giants and therefore not specific to environment or evolutionary stage.  
The observed abundance patterns are expected in material that has undergone proton captures at high temperatures -
enhancements in N, Na, and Al are accompanied by corresponding depletions in C, O, and Mg abundances\footnote{In the most widely accepted scenario, the fact that GCs display intrinsic variations in their light-element content points to different stellar generations, whereby a fraction of first generation stars polluted the interstellar matter from which second generations have born. However, a different model has been proposed by \cite{bastian13}.}. The C-N and Na-O anticorrelations display very different levels of enhancement/depletion across different systems suggesting that the pollution mechanisms and the nature of the polluters may actually vary from cluster to cluster. Observations of the Mg-Al anticorrelation (although not always present) imply a similar conclusion: the multiple populations are universal, but the progenitors are not. \cite{carretta09b} found that the extent of the Al enrichment  (and possible corresponding Mg depletion), is not the same in all GCs - it is related to a combination of cluster mass and metallicity. 
The \cite{carretta09b} result has been corroborated by \cite{meszaros15} for a different set of systems.  They confirmed that the spread in Al abundances increases significantly as cluster average metallicity decreases. They 
interpreted this as evidence that low-metallicity, intermediate-mass AGB polluters were more common in the metal-poor clusters. 
In addition to  AGB stars (\citealt{cottrell81}; \citealt{ventura01}), other candidates suggested as viable sources of internal pollution include fast-rotating massive stars
(FRMS, e.g., \citealt{decressin07}), massive binaries (\citealt{demink09}), or novae (\citealt{mz12}). Unfortunately, no one theory can 
satisfactorily account for the observed chemical patterns.

We have commenced a large observational survey with the scientific motivation to shed light on the nature of the internal polluters in GCs. Our focus has been on the element lithium which we consider to be one of the most powerful diagnostics of the progenitors responsible. 
The scientific context has been thoroughly discussed in our recent work (\citealt{dorazi14}, hereinafter Paper~{\sc i}), 
where we presented Li and Al abundances in GC giants of NGC~6218 (M 12) and NGC 5904 (M 5). In essence, Paper~{\sc i} is based on a straightforward nucleosynthetic prediction:  if the polluters produce no Li, then an anticorrelation should develop between Li and Al. This is because the second generation stars (Al-rich) will form from (almost) Li-free ejecta, whilst the primordial population are characterised by high Li abundances 
(appropriate to a value initially around the value of the Spite plateau, but modified by first dredge-up -FDU) and a modest Al content
(as field stars at the same cluster metallicity).
Furthermore, this anticorrelation would be erased if Li production occurs across the different stellar generations. Evidence for Li production in the progenitors was previously  detected in the GC M4 by \cite{dm10} and subsequently confirmed by other studies (e.g., \citealt{mucciarelli11}). In the two clusters analysed in Paper~{\sc i}, we determined that Li production in the progenitors is necessary to explain the class of stars that are both Li and Al rich. The conclusion favours AGB stars as responsible for internal pollution because their chemical yields can predict Li production via the so called Cameron \& Fowler (\citeyear{cf71}) mechanism. 
Curiously, the Li content measured in second-generation stars is exactly the same as the primordial population, requiring a kind of {\it ad hoc} fine tuning in the Li synthesis process. Current stellar theory does not allow for such Li production in other candidates. 

All three clusters, M12, M4 and M5  are of similar metallicity, with the last the most massive. In M5 we identified an abundance pattern not present in M4 and M12: the cluster hosts an extreme population of stars that are Li-deficient and Al-rich 
but which cannot be explained by a simple dilution model. 
Moreover, our comparison with previous studies yielded a correlation between the internal Li spread and the current cluster mass: the Li production seems less efficient in more massive GCs, whereas in smaller systems first-generation (FG) and second-generation (SG) stars share an almost identical Li content. We noted that small statistics and possible systematics may impact upon the speculative conclusions drawn from the comparison and  
thus we improve upon the sample size here.

In the present paper, we report the results from our study of Li and Al abundances for cluster giants in three more clusters, namely NGC~1904, ~NGC~2808, and NGC~362.
This study complements our previous work focussed on M12 and M5 and offers the largest Li survey in GC RGB stars available to date. 

\section{Observations, data reduction and analysis} \label{sec:obs}

\begin{figure*}
\includegraphics[width=2\columnwidth]{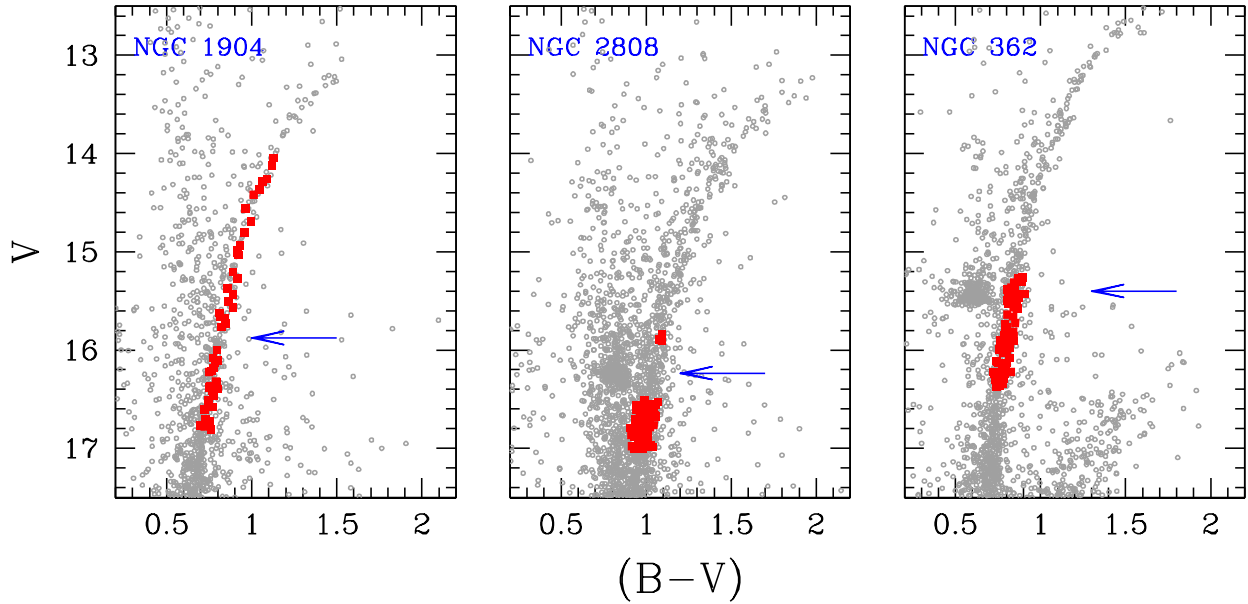}
\caption{Colour-magnitude diagram for the three globular clusters. The location of the RGB bump is marked (from \citealt{nataf13}). Stars analysed in the present study are emboldened.}\label{f:cmd}
\end{figure*}

Observations were carried out in visitor mode with FLAMES (\citealt{pasquini02}) mounted at UT2 of ESO-VLT on December 11-13 2014 
(Program 094.D-0363(A)). Spectra were collected with the scientific aim of inferring Li and Al abundances for large samples of RGB stars, below and above the RGB bump luminosity, in several GCs, as part of our dedicated survey (see Paper~{\sc i}). 
We utilised the HR15N high-resolution setup, covering simultaneously H$_\alpha$ and the Li doublet at 6708 \AA, which grants a nominal resolution of R=17,000 and a spectral coverage from 6470 to 6790 \AA. We selected our target stars based on photometric information published in \cite{momany04}, that is stars lying on the ridge line of the giant branch and with no companion brighter than 2 mag within 2$^{''}$. 
The colour-magnitude diagrams (BV photometry) for the three target clusters are shown in Fig.~\ref{f:cmd}, where the location of the bump is marked (from \citealt{nataf13}). 
The initial samples comprised 55, 106 and 78 giants in NGC~1904, NGC~2808, and NGC~362, respectively. 

\begin{figure*}
\includegraphics[width=2\columnwidth]{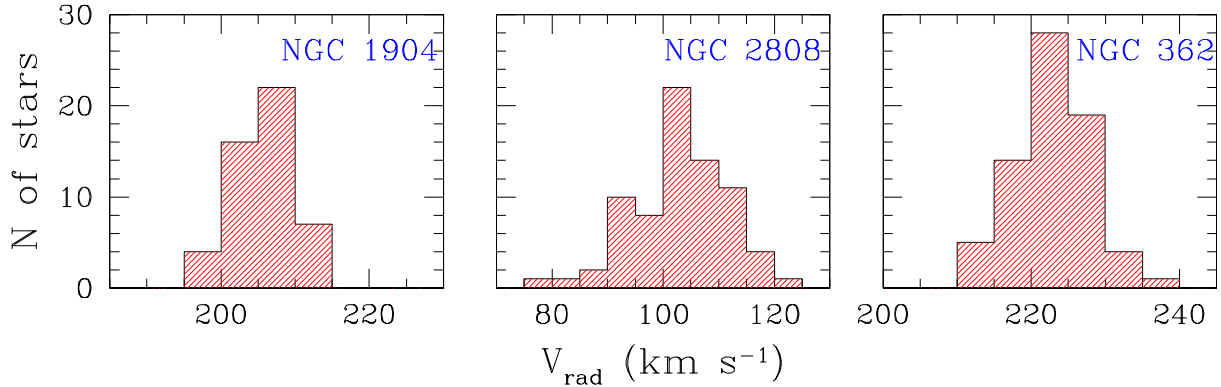}
\caption{Radial velocity distributions for our target stars.}\label{f:histo}
\end{figure*}

Data were reduced using the dedicated ESO pipeline (version 2.12.2, available at \url{http://www.eso.org/sci/software/pipelines/}), providing bias-subtracted, flat-fielded, optimal extracted and wavelength calibrated one dimensional spectra. 
Continuum normalisation, shifting to rest frame, and combination of multiple exposures were then performed using IRAF\footnote{IRAF is the Image Reduction and Analysis Facility, a general purpose software system for the reduction and analysis of astronomical data. IRAF is written and supported by National Optical Astronomy Observatories (NOAO) in Tucson, Arizona.} tasks.
After radial velocity computations, six stars in NGC~1904, 31 stars in NGC~2808, and six stars in NGC~362 turned out to be not members of their respective clusters and were thus discarded. We obtained average values of V$_{rad}$= 205.78$\pm$0.54 km~s$^{-1}$ (NGC~1904), V$_{rad}$= 102.79$\pm$0.97 km~s$^{-1}$ (NGC~2808), and V$_{rad}$= 222.95$\pm$0.61 km~s$^{-1}$ (NGC~362), which agree very well with measurements reported in Harris (\citeyear{harris96}, 2010 update), that is V$_{rad}$= 205.8$\pm$0.4 km~s$^{-1}$, V$_{rad}$= 101.6$\pm$0.7 km~s$^{-1}$, and V$_{rad}$= 223.5$\pm$0.5 km~s$^{-1}$, respectively. Our results are also in very good agreement with previous measurements by \citeauthor{carretta09a} (\citeyear{carretta09a}, \citeyear{carretta06}, \citeyear{carretta13}) of V$_{rad}$=205.60$\pm$0.48 for NGC~1904, V$_{rad}$=102.90$\pm$1.01 for NGC~2808 and V$_{rad}$=223.31$\pm$0.53 for NGC~362. The heliocentric radial velocity distributions for cluster members are shown in Figure~\ref{f:histo}.

As for Paper~{\sc i}, stellar parameters were derived in the following manner. Initial effective temperatures (\Teff) were calculated 
from ($V-K_s$) colours (with optical photometry retrieved from \citealt{momany04} and infra-red $K_s$ magnitudes from the 2MASS point source catalogue, \citealt{skru06}) and the calibration by \cite{alonso99}. To do this we assumed reddening and metallicity values collected in the Harris' catalogue, that is E(B-V)=0.01 and [Fe/H]=$-$1.60 for NGC~1904, E(B-V)=0.22 and [Fe/H]=$-$1.14 for NGC~2808, and E(B-V)=0.05 and 
[Fe/H]=$-$1.26 for NGC~362. As done in our previous studies, we converted E(B$-$V) values to E(V-K) adopting the relationship by \cite{cardelli89}. The final, adopted temperatures come from a relationship between those \Teff\ and $V$ magnitudes. 
Our samples consist of 22 RGBs in common with \cite{carretta09a} for NGC~1904 and 21 stars in NGC~362 already presented by \cite{carretta13}, 
the mean difference in temperature is $\Delta$\Teff=5.95$\pm$5.90 K and $\Delta$\Teff=$-$32.48$\pm$5.00 K. The larger difference for NGC~362 is probably related to the fact that in Carretta's study (\citeyear{carretta13}), \Teff\ values come from the relationship between K$_S$ magnitudes and VK colours, whereas we used the V magnitude (as done for NGC~1904 by \citealt{carretta09a}).

Surface gravities (log~$g$) were calculated assuming those \Teff, M$_{bol,\odot}$=4.75, the bolometric correction from \cite{alonso99}, a mass of M=0.85M$_\odot$, and distance moduli from Harris, by using the following equation:
$$\log\frac{g}{g_\odot}=\log\frac{M}{M_\odot}-\log\frac{L}{L_\odot}+4\log\frac{T_{\rm eff}}{T_{\rm eff,\odot}}$$.

%
%\centering

Finally, microturbulence velocities were computed with the relationship by \cite{gratton96}, that is $\xi =2.22 - 0.322 \times \log \ g$

\begin{figure*}                                                      
\includegraphics[width=2\columnwidth]{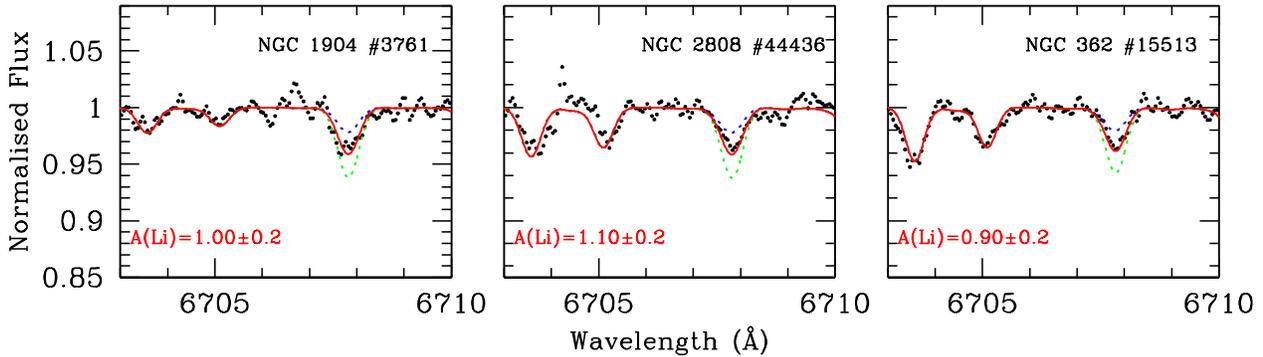}                       
\caption{Examples of spectral synthesis for one star in each target cluster.}\label{f:synth}
\end{figure*}

Similarly to Paper~{\sc i}, we gained abundances for Li and Al from the strong doublets at 6708 \AA~and 6696/6698 \AA, respectively. 
Due to the occurence of cosmic rays on the spectral features under consideration and/or to lower S/N ratios, we could obtain Li abundances for 
47 stars in NGC~1904, 68 stars in NGC~2808, and 67 stars in NGC~362. The Al content has been inferred for 48, 65, and 66 cluster members, respectively. 
We performed spectral synthesis calculations using the local thermodynamical equilibrium (LTE) code 
{\sc MOOG} by C. Sneden (\citealt{sneden73}, 2014 version) and the \cite{ck04} grids of model 
atmosphere\footnote{Available at \url{http://kurucz.harvard.edu/grids.html}}, with $\alpha-$enhancement (+0.4 dex) and no convective overshooting. 
An example of spectral synthesis for the Li feature is displayed in Figure~\ref{f:synth}.
We finally applied NLTE corrections to our Li abundances following prescriptions by \cite{lind09}.
%
%\centering                                                           

Random uncertainties were calculated as described in Paper~{\sc i}, to which we refer the reader for an extensive discussion. 
Here we briefly recall that (internal) error measurements include two kind of contributions: uncertainties related to the best-fit determination and errors due to the adopted set of atmospheric parameters. Given their independence, they can be added in quadrature in order to derive the final internal uncertainty on Li and Al abundances (see Paper~{\sc i} for details). Typical values range from 0.08 - 0.12 for Li and 0.10 - 0.15 for Al.

\section{Results and discussion}\label{sec:results}
Results of our analysis are reported in Table~\ref{t:param}: optical and {\sc 2MASS} K$_s$ photometry, heliocentric radial velocity, S/N ratios as calculated around the Li~{\sc i} doublet and final stellar parameters (effective temperature, gravity and microturbulence) are listed for each target star in the three clusters
(the full table is available in the online version). Both the LTE and NLTE Li abundances are presented in 
Columns 10 and 11, respectively, while [Al/Fe] ratios are given in Column 12 
(asterisks denote upper limits in their respective measurements).  

\subsection {The trend of lithium with luminosity}

\begin{figure*}
\includegraphics[width=2\columnwidth]{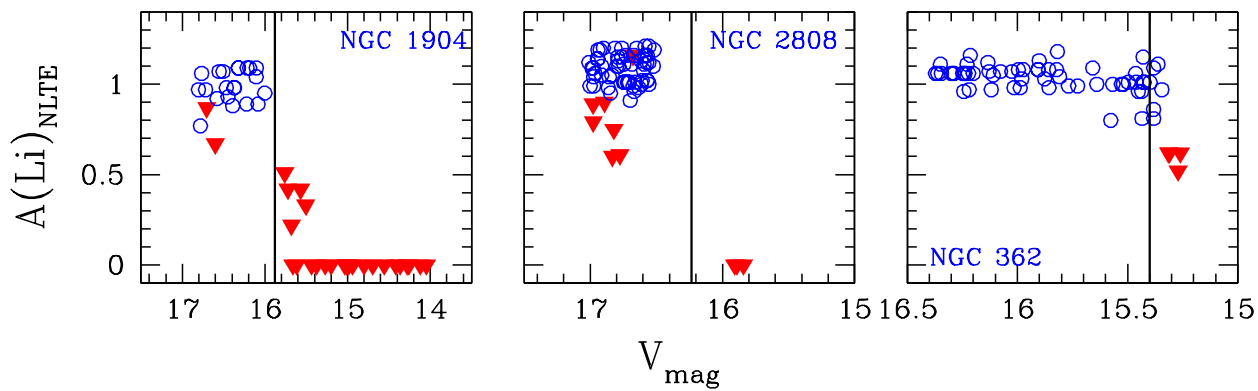}
\caption{NLTE Li abundances as a function of V magnitudes. The vertical bar marks the RGB bump. Empty circles denote measurements and upside-down triangles are for upper limits}\label{f:limag}
\end{figure*}

We plot Li as a function of V magnitude in Figure~\ref{f:limag} for NGC~1904 (left-hand panel), NGC~2808 (middle panel), and NGC~362 (right panel). The locations of the RGB bump in each cluster are taken from \cite{nataf13} and are marked by vertical lines at V$_{bump}$=15.88, 16.24 and 15.40, respectively. 
As expected from standard stellar evolution, stars that have already experienced the FDU are depleted in Li
by factors 15-20 with respect to the Spite value (thus if we assume that this is the pristine Li abundance, with respect to their original content, although some further extra mechanisms such as atomic diffusion might have played a minor role).
We determined mean lithium abundances of 
A(Li)$_{\rm NLTE}$= 0.97 $\pm$ 0.02 (rms=0.11, 23 giants) for NGC~1904, A(Li)$_{\rm NLTE}$= 1.06 $\pm$ 0.02
(rms=0.13, 65 stars) for NGC~2808 and A(Li)$_{\rm NLTE}$=1.02 $\pm$ 0.01 (rms=0.09, 57 stars) for NGC~362 by considering only those stars fainter than the bump luminosity. All three clusters share the same mean Li abundance once observational uncertainties are taken into account, and furthermore, they agree with the Li abundances determined for every other GC giant currently in the literature\footnote{A very small fraction, below the 1\%, of Li-rich giants are also detected in GCs (see e.g., our recent work, \citealt{dorazi15}, and references therein). The nature of these objects is however related to a peculiar evolution and hence its discussion is not relevant to the present work.}. 
 If we apply NLTE corrections to data published in Paper~{\sc i}, we obtain 
A(Li)$_{\rm  NLTE}$=1.07$\pm$0.01 for NGC~6218 and 1.02$\pm$0.01 for NGC~5904 (average abundances under the local thermodynamical equilibrium approximation were 0.98$\pm$0.01 and 0.93$\pm$0.01, respectively). We stress that giants for these five GCs are analysed in a strictly homogeneous way (same code, line-lists, model atmospheres and techniques) and consequently systematics do not impact on our conclusions. We find also that our derived Li abundances are similar to measurements by \cite{mucciarelli14} in the massive GC M54.
M54 is located in the centre of the Sagittarius dwarf spheroidal galaxy (\citealt{ibata94}) and is known to exhibit a spread in metallicity and $s$-process elements (e.g., \citealt{carretta10a}). Mucciarelli and collaborators derived a mean value of A(Li)=0.93$\pm$ 0.11 (or 1.01, depending on the adopted NLTE corrections).
Thus our results seem to indicate that GC giants (before they reach the RGB bump) share the same primordial Li abundances, independent of cluster's mass, metallicity, horizontal branch morphology, or location in the Galaxy (disk vs halo) or beyond (Milky Way $vs$ dwarf Spheroidal GCs).
We note that it is also possible that the stars are born with a different Li content but FDU or some rapid pre main-sequence process may operate in such a way that they eventually end up with the same Li abundances. Stellar models currently do not lend support to the FDU scenario whilst PMS depletion may require some yet to be included physical process.  
We summarise the GC global structural parameters along with 
Li abundances in Table~\ref{t:clusters}, where we report values published in Harris's catalogue for metallicity ([Fe/H]), absolute cluster magnitude (a proxy for the current mass), King-model central concentration (a c denotes core-collapsed cluster). We include also the horizontal branch ratios from \cite{gratton10}, the interquartile range of Na-O distributions as well as fractions of primordial (P), intermediate (I) and 
extreme (E) populations\footnote{The cluster populations are defined according to their O and Na abundances: P stars are those having [Na/Fe]$\leq$[Na/Fe]$_{\rm min}$ + 4$\sigma$, whereas second-generation stars are classified as intermediate or extreme if [O/Na] is larger or smaller than $-$0.9 dex, respectively.} from \cite{carretta10b} and  \citeauthor{carretta13} (\citeyear{carretta13}, see also following discussion).

Once stars evolve past the bump in the luminosity function, their Li content is efficiently depleted due to the onset of some kind of extra mixing processes: the H-burning shell erases the molecular weight discontinuity (the so called ``$\mu$-barrier'') established by first dredge-up and deep mixing occurs. The nature of this kind of extra mixing is one of the hottest topic in current stellar evolution theory and different mechanisms have been suggested such as e.g., ``thermohaline mixing" (\citealt{eggleton06}) or magnetic buoyancy (\citealt{busso07}).
For clusters NGC~1904 and NGC~362, where we well sample this region of the HR diagram,  the location of the photometric and spectroscopic RGB bumps match very well: the onset of lithium depletion coincides with the bump magnitude. In the case of   
 NGC~2808, their is a dearth of observations at this magnitude due to an observational bias in our target selection. Our sampling is somewhat discontinuous but it is clear that stars beyond the bump luminosity possess undetectable levels of Li. 

\begin{table*}
%\centering
\caption{Photometry, stellar parameters, kinematics and chemical information for our sample stars. The complete table is made available on line.}\label{t:param}
\hspace*{-0.9cm}
\tabcolsep=0.13cm
\begin{tabular}{lccccccccccccr}
  \hline \hline
Star ID   & RA & DEC  & $B$   & $V$     &  K$_s$ & V$_{\rm rad}$  &  S/N   & \Teff &  $\log \ g$     & $\xi$ & A(Li)$_{\rm LTE}$ & A(Li)$_{\rm NLTE}$ & [Al/Fe] \\
          & (hh:mm:ss) & ($^o$:':'')      & (mag)  & (mag)   &  (mag) &  (km s$^{-1}$) & (@6708\AA) & (K) &  (cm s$^{-2}$) & (km s$^{-1}$) & dex & dex & dex\\
  \hline
1904-4196 & 05:23:58.912 & $-$24:30:02.012 & 16.798 & 16.003  & 13.698 &   209.82     &  188    & 4841   &  2.20    & 1.51  &    0.85  &  0.95    &   -0.20\\   
1904-8925 & 05:24:17.168 & $-$24:31:24.698 & 16.858 & 16.083  & 13.896 &   203.43     &  175    & 4858   &  2.24    & 1.50  &    0.80  &  0.89    &    0.60\\   
1904-6388 & 05:24:11.917 & $-$24:32:37.674 & 16.900 & 16.105  & 13.916 &   202.70     &  180    & 4863   &  2.25    & 1.50  &    1.00  &  1.09    &   -0.15\\   
1904-7558 & 05:24:13.986 & $-$24:32:01.426 & 16.911 & 16.107  & 13.913 &   204.79     &  160    & 4863   &  2.25    & 1.50  &    0.95  &  1.04    &   -0.10\\   
1904-8251 & 05:24:17.435 & $-$24:31:42.530 & 16.912 & 16.131  & 13.880 &   208.93     &  170    & 4868   &  2.26    & 1.49  &    ....  & ....     &   ~0.60\\
\hline
\end{tabular}
\label{tab:param}
\end{table*}  

\begin{table*}
%\centering
\caption{Structural properties and Li abundances (for giants below the bump luminosity) for clusters from our survey}\label{t:clusters}
\hspace*{-0.9cm}
\tabcolsep=0.13cm
\begin{tabular}{lcccccccccc}
  \hline \hline
Cluster  & [Fe/H]  &  M$_V$  & c         &   HBR       & IQR[Na/O] & P & I & E          &  $<$A(Li)$>$ & rms\\
         &  (dex)  &   (mag) &          &              &           & (\%) & (\%) & (\%) & (dex) & (dex) \\
\hline
         &         &         &          &              &           &    &    & & \\
NGC 1904 & $-$1.60 & $-$7.86 & ~1.70c	&      ~~~0.89 &  0.759    & 40 & 50   & 	10   & 0.97 & 0.11	\\
NGC 6218 & $-$1.37 & $-$7.31 & ~1.34	&      ~~~0.97 &  0.863    & 24 & 73   & 	3    & 1.07 & 0.06	\\
NGC 5904 & $-$1.29 & $-$8.81 & ~1.73	&      ~~~0.31 &  0.741    & 27 & 66   & 	7    & 1.02 & 0.11	\\
NGC 362  & $-$1.26 & $-$8.43 & 1.76c	&      $-$0.87 &  0.644    & 22 & 75   & 	3    & 1.02 & 0.09	\\
NGC 2808 & $-$1.14 & $-$9.39 & ~1.56	&      $-$0.49 &  0.999    & 50 & 32   &        18   & 1.06 & 0.13	\\
         &         &         &          &              &           &   &    &   &\\
\hline\hline
\end{tabular}
\end{table*}

\subsection{Aluminium abundances}\label{sec:aluminium}
\begin{figure*}
\includegraphics[width=2\columnwidth]{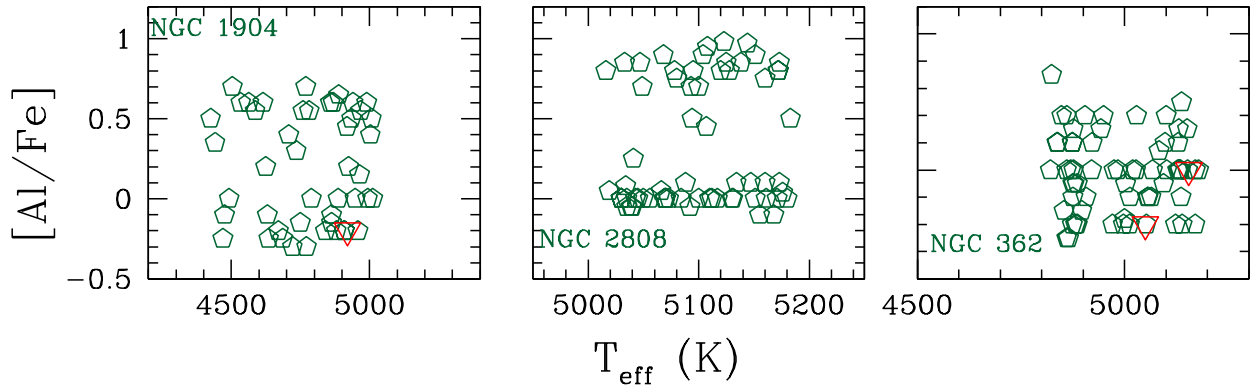}
\caption{[Al/Fe] ratios $versus$ effective temperatures (\Teff) for the whole sample. As for Figure~\ref{f:livmag} upside-down triangles denote upper limits}\label{f:alteff}
\end{figure*}

\begin{figure*}
\includegraphics[width=2\columnwidth]{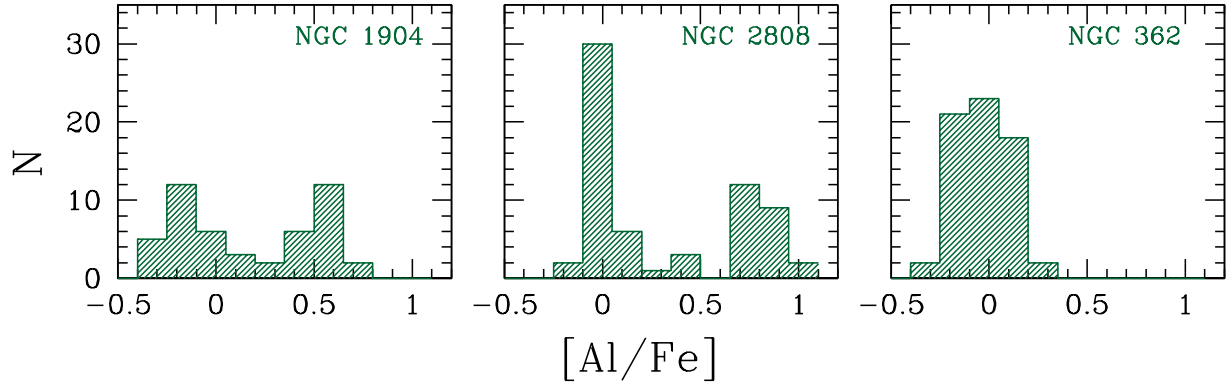}
\caption{Histograms of Al abundances for all giants under consideration}\label{f:al_histo}
\end{figure*}

We determined Al for 48, 65, and 66 giants (below and above the RGB bump) in NGC~1904, NGC~2808, and NGC~362, respectively (179 GC members in total). The [Al/Fe] ratios are shown as a function of \Teff\  for our target stars in Figure~\ref{f:alteff}; no obvious trend is evident from the data, corroborating the reliability of our spectral analysis.
In NGC~1904, we found an average value of [Al/Fe]=0.19$\pm$0.05 (rms=0.35), with abundances ranging from $-$0.30 to 0.70 dex, corresponding to a variation of a factor of 10 within this cluster. Our Al abundance determinations for NGC~1904 are offset by approximately 0.45 dex to the values determined by \cite{carretta09b}, i.e., [Al/Fe]=0.64$\pm$0.15, rms=0.41.
We have no stars in common with their sample so a direct comparison aimed at revealing the cause of the discrepancy in not possible.  
We recall that Carretta's work exploited high-resolution UVES spectra and was then focussed on brighter giants, thus NLTE effects might contribute to the difference. Their sample was also smaller than ours with only one of their eight RGB stars labelled `primordial' ([Al/Fe]=$-$0.216). Their remaining seven stars display different levels of enhancement and are considered second generation members. It is interesting to note that the \citet{carretta09b} results imply that  the relative distribution of different stellar populations determined from Al enhancement does not reproduce the distribution inferred from Na and O abundances. \cite{carretta09a}
obtained fractions of FG=40\% and SG=60\% (50\% I + 10\% E populations) which better reflects our Al distributions (see Figure~\ref{f:al_histo}, where the two peaks in Al abundances have almost the same size). It is tempting to speculate that limited sample size in the \cite{carretta09b} survey may skew the statistics; however the internal variation in the Al content agrees very well between the two studies (with Carretta's study providing a lower limit to the Al spread of $\gtrsim$ 0.85 dex).

For NGC~2808 we derived a mean Al abundance of [Al/Fe]=0.32$\pm$0.05 (rms=0.39), which is consistent with Carretta's (\citeyear{carretta09b}) determination of [Al/Fe]=0.40$\pm$0.14 (rms=0.49) and with the recent analysis by \cite{carretta14b}, i.e., [Al/Fe]=0.46$\pm$0.09 (rms=0.47). 
We detected an internal spread of more than 1 dex (-0.1 to 0.98 dex) agreeing with previous 
findings by \cite{bragaglia10} based on unevolved stars.
They published the first measurement of Al abundances for main-sequence stars in NGC~2808 using {\sc $X$-shooter}@VLT spectra, deriving [Al/Fe]=$-$0.2 for the red main-sequence star (i.e., it belongs to the primordial population of the cluster) and [Al/Fe]=+1.1 for the (blue) second-generation dwarf.
We also confirm the finding by \cite{carretta14b} that the Al distribution is clearly trimodal in NGC~2808 (see Figure~\ref{f:al_histo}). A possible hint for a double-peak trend in the Al content seems to be present for NGC~1904, but the pattern is less clear (we also note that due to the lower resolution and signal-to-noise (S/N) ratios of our spectra, observational uncertainties might be responsible for smearing off the distribution).

The spread of [Al/Fe] in NGC~362 spans 0.6 dex (range $-$0.25 dex to +0.35 dex), with an average abundance of 
[Al/Fe]=$-$0.02$\pm$0.02 (rms = 0.14). We find a $\sim$ 0.25 dex  offset compared to the value reported by 
 by \cite{carretta13}, namely [Al/Fe]=0.24$\pm$0.05, rms=0.19. As we speculated for NGC~1904, the differences between the two studies might be related to sample size, small differences in temperature scales and/or NLTE effects and we are once again in the unfortunate position where our surveys contained no stars in common. There is yet again good agreement between the observed internal spreads with  Carretta et al. quoting a peak-to-peak variation of 0.56 dex.

Although for two of our clusters we derive different mean [Al/Fe] abundances to those previously published in the literature, 
our discussions and conclusions remain unaffected. For our scientific analysis, zero-point offsets will not impact 
upon the fact the large spread in Al is evidence for multiple populations. In Figure 7 we plot our [Al/Fe] as a function of [Na/Fe] and [O/Fe] as determined by \cite{carretta09a} and \cite{carretta13} for NGC~1904 and NGC~362 (we have no stars in common for NGC~2808). 
We still recover the pattern of a very extended positive correlation between Na and Al for NGC~1904 as well as a significant O-Al anticorrelation. The Pearson correlation coefficient is r=0.84 (18 degrees of freedom, significance level more than 99.9\%) and r=$-$0.83 (16 degrees of freedom, significance level more than 99.9\%), respectively for Na and O. 
NGC~362 exhibits a less pronounced Al variation (see also \citealt{carretta13}) with respect to the more massive GC NGC~2808 or the more metal-deficient cluster NGC~1904. However, statistically meaningful (anti)correlations with O and Na do appear, with Pearson correlation coefficients of r=0.83 for Na and r=$-$0.71 for O. The probabilities that those correlations happened by chance are also in this case less than 0.1 \%. We thus corroborate previous results by \cite{shetrone00} and \cite{carretta13} who compared the second-parameter pairs NGC~288 and NGC~362. The light-element variations in NGC~362 are more extreme (a factor of two in [Al/Fe]) probably owing to its larger mass (they have the same metallicity but NGC~362 is five times larger than NGC~288).

Our findings confirm that the extent of internal variations in proton-capture elements, and in particular the presence of species requiring very high burning temperatures such as e.g., aluminium, depends upon a combination of the cluster's mass and metallicity. 
This result was first reported by \cite{carretta09b} and later on verified in other clusters by \cite{meszaros15}.

\begin{figure*}
\includegraphics[width=1.5\columnwidth]{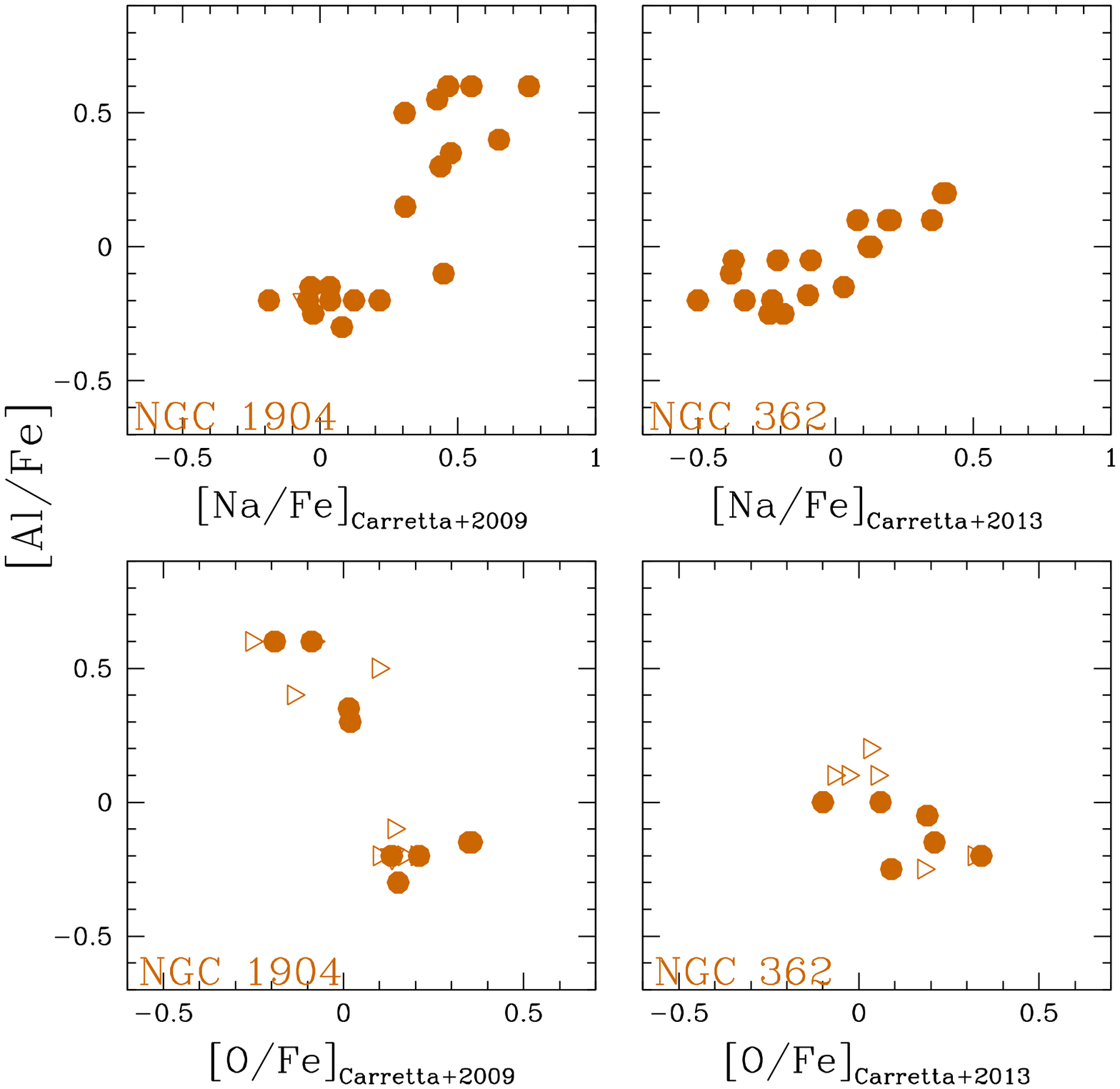}
\caption{Run of [Al/Fe] with [Na/Fe] and [O/Fe] for stars in common with \citet{carretta09a}. Filled circles and empty triangles are for measurements and upper limits, respectively. No stars are in common for cluster NGC~2808.}\label{f:alnao}
\end{figure*}

\subsection{Lithium abundances and the multiple population framework}\label{sec:spreads}

\begin{figure*}
\includegraphics[width=2\columnwidth]{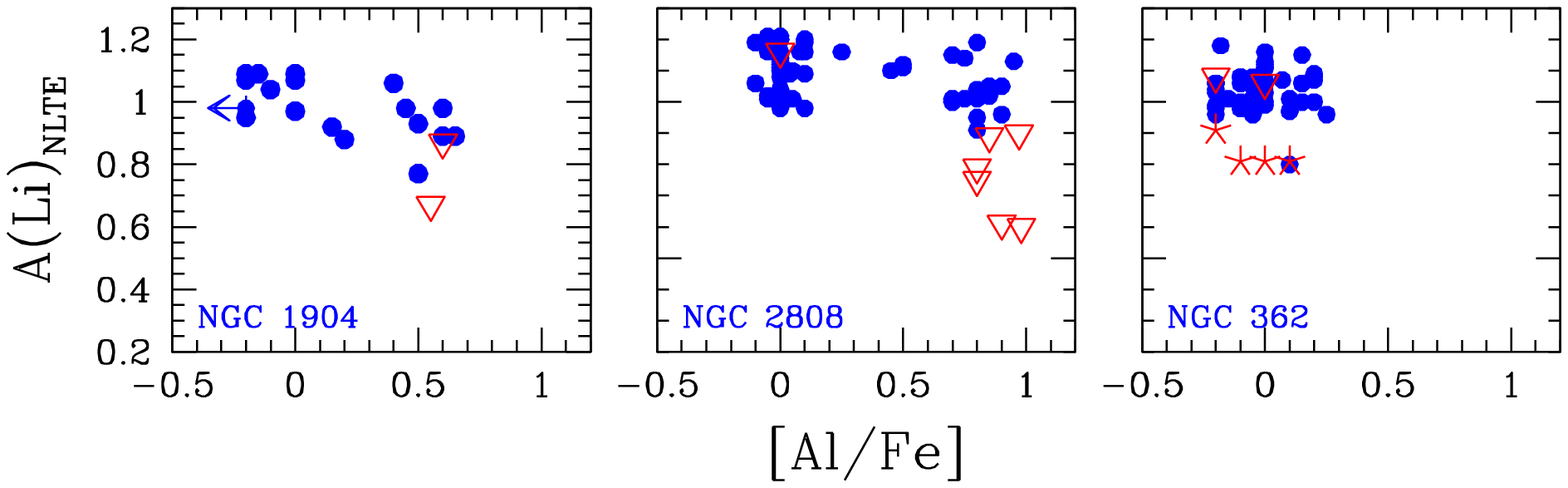}
\caption{NLTE Li abundances as a function of [Al/Fe] ratios for stars fainter than the bump luminosity (see text for discussion).}\label{fig:lial}
\end{figure*}

The determination of Li abundances in conjunction with p-capture elements (e.g., Na, O) allows us to put strong observational constraints on the internal pollution mechanisms and their corresponding stellar source.  
All three clusters analysed in this study exhibit significant internal variations in their Al content (see Section~\ref{sec:aluminium}), implying that temperatures in excess of 70$\times$10$^6$ K have been reached within the first generation stars that provided the enriched ejecta to give birth to subsequent stellar generations. 
It is quite reasonable to expect that at those high temperature all Li is destroyed (Li starts burning at only 2.5$\times$10$^6$ K) and that their second generation stars are born of Li-free ejecta (though a mix with some pristine material has to be postulated regardless of the polluters' type, e.g., \citealt{prantzos07}; \citealt{dercole08}, \citeyear{dercole11}; \citealt{conroy11}). 
It follows that the primordial GC population should possess lithium at the  Spite plateau level whereas the  second-generation stars should demonstrate lower Li abundances (the extreme population should have virtually no Li if we assume they come from pure ejecta).
Our observations can test this scenario because our analysis would detect {\it conventional} (anti)correlations between Li and proton-capture elements (that is the positive correlations Li-O, Li-C, and Li-Mg along with anticorrelations between Li and Na, N, Al). 
If Li production took place within the polluters the anticorrelations are substantially erased. 

During our previous investigations, we found that Li must have been produced in the progenitors that polluted the 
GCs M4 (\citealt{dm10}) and  M12 (Paper~{\sc i}): despite having their O content reduced by factors 3-4, second-generation stars in these clusters are still Li rich. Similar findings have been also obtained by \cite{mucciarelli14} for the extra-galactic GC M54, where Li 
abundances in the first and second generation stars are essentially the same.
In Figure~\ref{fig:lial} we present our Li abundances as a function of [Al/Fe] ratios (for stars below the bump luminosity).
In NGC~1904 we detected evidence for a Li-Al anticorrelation, with the Pearson correlation coefficient r=$-$0.67 (significant at more than 99.9\% level). However, the observed pattern reproduces the trend we discovered in the GC NGC~5904 (M5) (see Paper~{\sc i}): 
both first and intermediate generation stars contain the same Li abundance suggesting internal polluters able to produce Li, such as intermediate-mass AGB stars. 
The clusters M5 and NGC~1904 differ from M4 and M12 in that they host an extreme population that exhibit a Li-deficient pattern. Since we could only derive upper limits, we are not able to ascertain whether this extreme population is totally Li-free or a small amount of Li has been produced. 
The other key feature emerging from our study is that a simple dilution model would fail in reproducing the observed pattern, by assuming that the Intermediate population is a simple mixture of primordial and extreme population (as found for M5, see Paper~{\sc i} for further details).

The massive cluster NGC~2808 displays a very similar pattern in terms of Li and Al abundances to M5 and NGC~1904: the correlation coefficient between Li and Al is r=$-$0.57 (62 degrees of freedom, significance level at more that 99.9\%). Analogous to NGC~1904, the intermediate population is still significantly Li-rich (despite being enhanced in Al), whereas the extreme population shows evidence of Li depletion (to different levels). It is noteworthy that in this case the extreme component is more numerous than for NGC~1904 and remains in agreement with the relative fractions of different stellar populations derived by \cite{carretta09a} according to Na and O abundances. The fraction of E stars in NGC~2808 is almost twice that of NGC~1904, probably because of its large mass. Thus, in both these GCs we require a Li production between the first and the intermediate second-generation stars (and hence AGB stars seem to provide us with a culprit), whereas we are not able to discard any of the candidate polluters to account for the observed pattern in extreme stars. They indeed could be either fast-rotating massive stars or asymptotic giant branch stars that did not efficiently produce Li (likely due to a combination of their mass and metallicity regimes).

Conversely, NGC~362 displays a similar behaviour to that which we have observed in M12 and M4. There is no Li difference across the stellar generations. A Pearson correlation coefficient here of r=$-$0.04 is not statistically meaningful, implying no correlation between [Al/Fe] and A(Li) exists. In Figure~\ref{fig:lial} there are a handful of stars showing a slightly lower Li abundance (marked as asterisks in the plot). However all these giants are located very close to the bump luminosity, namely they are all within 0.3 mag from the bump magnitude quoted by \cite{nataf13}. This implies that their lower Li content might be actually due to evolutionary effects and not necessarily related to the multiple populations. 
Alternatively, they could be binary stars, showing a Li-depleted nature as a consequence of their binary evolution.
However, an offset between our photometry and the study by Nataf and collaborators is present and we indeed detect the RGB bump at 0.45 mag fainter than theirs  (V$_{bump}$=15.46). Had we adopted this value for the bump luminosity, we could have discarded those stars from the present discussion. As a consequence, our results tentatively indicate that significant Li production has occurred in NGC~362  with  AGB stars seemingly responsible.  

In Paper~{\sc i}  we reported the  detection of a negative correlation between the absolute cluster magnitude (a proxy for the current mass) and the internal Li variation (as given by the peak-to-peak variation, see Figure~14 in that paper). We interpreted this 
as evidence of different efficiency in the Li synthesis mechanisms within different clusters, with 
smaller GCs exhibiting a more conspicuous level of Li production (that makes first and second generations indistinguishable in terms of their Li content). The relationship was proposed with the caveat that we were unable to discern the role of metallicity in such a limited sample of clusters. The results presented here do in fact suggest that metallicity plays a role in determining the internal spread in lithium. NGC~362 is much less massive than NGC~2808 and has a metal content more than a factor of two higher than NGC~1904 which indicates that Li production is less efficient in more massive and more metal-poor systems. \cite{carretta09b} reached the same conclusion about the level of Al spread in their cluster sample. 

Finally, it would be tempting to speculate that the reason for the high Li content in second-generation stars might be related to their He-rich nature. It is possible that FDU is less efficient in these stars resulting in less Li depletion. 
We ran stellar models (Angelou et al., in prep) with calculations for FG and an extreme population stars which indicate differences at the level of 
A(Li)$\sim$ 0.1 dex owing to the respective convective envelope penetration. These differences lie within the  
observational uncertainty and are therefore indistinguishable.
Nevertheless, if the He abundance is responsible we should expect high Li in E stars compared to I stars as E stars are thought to contain a very high level of He enrichment (up to Y$\sim$0.4 suggested for NGC~2808). We demonstrated that, whenever present, the E component is actually more Li-poor (possibly even Li-free) than the intermediate population.

\section{Summary and Concluding remarks}\label{sec:summary}
The aim of our survey has been to probe the stellar sources of internal pollution in GCs.
We have determined abundances for Li and Al in a sample of
roughly 180 giant stars in clusters NGC~1904, NGC~2808, and NGC~362.
This investigation complements our previous work which focussed on NGC~6218
(M12) and NGC5904 (M5) and provides the community with the largest,
homogeneous database of GC lithium abundances available in the literature to date.
The robustness of lithium as diagnostics of the enrichment mechanisms in
GCs, and its ability to give us a unique constraint on the stellar type,
is based on a straightforward nucleosynthetic prediction: if
there was no Li production in the polluters, then we should detect
positive correlations between Li and C,O, and Mg and hence
anticorrelations between Li-N, Li-Na, Li-Al. Per contra, if Li production 
has occurred in the polluters then  the common (anti)correlations are erased -- the first- and second-generation stars
are inclined to exhibit similar Li abundances. This has proven to
be the case in clusters such as e.g., M4 (\citealt{dm10};
\citealt{mucciarelli11}; \citealt{vg11}; \citealt{monaco12}), M12
(Paper~{\sc i}), and M54 (\citealt{mucciarelli14}). 
Li production appears to have also occurred in M5 
(as we report in
Paper~{\sc i}), in NGC~6397 (\citealt{lind09}), and in NGC~6752 (e.g.,
\citealt{shen10}), but these GCs also host a distinct extreme population that is somehow
Li-depleted (or virtually with no Li, since only upper limits are
available). Interestingly, the chemical behaviour of intermediate stars
in these GCs are {\it not} the outcome of a simple mixture  of pristine
material and the composition measured in their extreme populations. The result
implies that at least two kind of pollution events (thus two different
kind of stellar polluters) are required to account for the observational
constraints. In the current study we have identified that NGC~362 demonstrates 
a similar pattern to that found in M4 and M12, with no extreme population
present and the first and second generation stars exhibiting the same Li
abundances (within the observational uncertainties). Conversely, we detected
a significant Li-Al anticorrelation in clusters NGC~1904 and NGC~2808,
driven by the presence of the E population, which are
Al-rich and Li-deficient. To account for the Li-rich intermediate populations in these clusters we 
need to postulate Li production between first- and
intermediate second-generation stars favouring AGB stars as the internal polluters. 
We are unable to conclude if Li replenishment took
place for the extreme cluster component and therefore we cannot discard any of the other polluter candidates as
progenitors to these stars. 
Finally, our findings seem to suggest that the internal Li
scatter and effective Li production is related to a combination of a
cluster's mass and metallicity with the relatively low mass and/or metal-rich clusters more efficient as far as this production is concerned.

\section*{Acknowledgments}

This work made extensive use of the SIMBAD, Vizier, and NASA ADS databases. 
We acknowledge partial support by the Australian Research Council (ARC), PRIN INAF 2011 ``Multiple populations in globular clusters: their role in the Galaxy assembly'' (PI E. Carretta) and PRIN MIUR 2010-2011 ``The Chemical and Dynamical Evolution of the Milky Way and Local Group Galaxies” 
(prot. 2010LY5N2T, PI F. Matteucci). The research leading to the presented results has received funding from the European Research Council under the European Community's Seventh Framework Programme (FP7/2007-2013) / ERC grant agreement no 338251 (StellarAges). 

\bibliographystyle{mn2e}

%bsp

\label{lastpage}

\end{document}